\documentclass[onecollarge,natbib]{svjour2}
\bibpunct{[}{]}{;}{n}{}{,} 
\smartqed  
\usepackage{color}
\usepackage{graphicx}

\def\la{\langle}
\def\ra{\rangle}

\def\lam{\lambda}

\def\be{\begin{equation}}
\def\ee{\end{equation}}
\def\bea{\begin{eqnarray}}
\def\eea{\end{eqnarray}}
\journalname{Few-Body Systems}
\begin{document}

\title{Self-consistent covariant description of twist-3 distribution amplitude of a
pseudoscalar meson in the light-front quark model
\thanks{This work was supported in part by the Korean Research Foundation
Grant funded by the Korean Government (KRF-2010-0009019).}}


\author{Ho-Meoyng Choi   \and
        Chueng-Ryong Ji 
}


\institute{H.-M. Choi \at
              Department of of Physics, Teachers College, Kyungpook National University,
              Daegu, Korea 702-701 \\
              \email{homyoung@knu.ac.kr}           
           \and
           C.-R. Ji \at
            Department of Physics, North Carolina State University, Raleigh, NC 27695-8202 \\
            \email{crji@ncsu.edu}
}

\date{Received: 15 September 2014 / Accepted: 11 November 2014}

\maketitle

\begin{abstract}
We discuss the light-front zero-mode issue in the light-front quark model prediction
of the twist-3 distribution amplitude of a pseudoscalar
meson from the perspective of the vacuum fluctuation
consistent with the chiral symmetry of QCD.
\keywords{Twist-3 distribution amplitude \and Light-front quark model \and Light-front zero-mode}
\end{abstract}

\section{Introduction}
\label{intro}
The meson light-cone distribution amplitudes (DAs) are important elements in applying QCD to hard
exclusive processes via the factorization theorem~\cite{LB80,ER80}.
In studying the exclusive processes, it is
expedient to arrange the meson DAs by different twist structures.
The leading-twist DA describes the distribution of valence quarks
in terms of the longitudinal momentum fractions inside the meson and
provides the main contribution to the exclusive processes.
The higher-twist DAs depict either contributions
from transverse motion of constituents in the leading-twist components
or from the higher Fock-states such as quark-antiquark pair terms.
Although the higher-twist DAs are suppressed by a factor of $1/Q^2$ in the large
momentum transfer $Q^2$ region compared to the leading twist-2 DAs, their contributions may
still be important as nonperturbative inputs together with the leading twist-2 DAs
for certain exclusive processes.

Perhaps one of the most convenient and intuitive framework in studying the hadronic DAs is
based upon the light-front dynamics (LFD).
Particularly, the rational energy-momentum dispersion relation of LFD yields the sign correlation between the LF energy $k^-(=k^0-k^3)$ and the LF longitudinal momentum $k^+(= k^0 + k^3)$ and leads to the suppression of vacuum fluctuations.
Despite these advantages in LFD, however, the zero-mode~($k^+ = 0$) complication in the
matrix element has been noticed not only for some electroweak form factors between mesons but also for the vector meson decay constant. Especially, meson decay constants are the lowest moments
of the meson light-cone DAs.
In our recent analysis of the vector meson decay constant and chirality-even
twist-2 and twist-3 quark DAs of the vector meson~\cite{CJ_V14}, we found that the zero mode
depends on the hadronic models with different choice of the LF wave functions.
To prove this point, we utilized two models, i.e. exactly solvable manifestly covariant
Bethe-Salpeter (BS) model with the multipole ansatz
for the $q\bar{q}$ bound-state vertex function
and the light-front quark model (LFQM)
with the more phenomenologically accessible Gaussian type
radial wave function. The LFQM using the standard light-front (SLF) approach has been quite successful in describing various static and non-static properties of hadrons. However,
as the SLF approach within the LFQM by itself is not amenable to determine the zero-mode contribution,
we utilized the manifestly covariant BS model to check the existence (or absence) of the zero mode.
Linking the covariant BS model to the standard LFQM, we found the correspondence relation between the two models to give a self-consistent covariant description of the vector meson decay constant within the LFQM. Remarkably, we found in~\cite{CJ_V14} that the nonvanishing zero-mode contributions as well as the instantaneous ones to the vector meson decay amplitude appeared in the covariant BS model vanish exactly in the LFQM with the Gaussian wave function.

The purpose of this work is to extend our previous work to analyze the decay amplitude related with twist-3 DAs of a pseudoscalar meson within the LFQM.
The twist-2 DA of a pseudoscalar meson has been analyzed in our previous work of LFQM~\cite{CJ_DA}.
While there are two independent twist-3 two-particle DAs of a pseudoscalar meson, namely,
$\phi^{\cal P}_{3;M}$
and $\phi^\sigma_{3;M}$
corresponding to pseudoscalar and tensor channels of a meson ($M$), respectively,
we shall study
$\phi^{\cal P}_{3;M}$
in the present work.

The
$\phi^{\cal P}_{3;M}$
is defined in terms of the
following matrix elements of gauge invariant nonlocal operators at light-like
separation~:
\be\label{Deq:1}
\la 0|{\bar q}(z)i\gamma_5 q(-z)|M(P)\ra
= f_M \mu_M \int^1_0 dx e^{i\zeta P\cdot z}
 \phi^{\cal P}_{3;M}(x),
\ee
where $z^2=0$ and $P$ is the four-momentum of the meson ($P^2=m^2_M$) and the integration variable $x$ corresponds to the longitudinal momentum fraction
carried by the quark and $\zeta =2x -1$ for the short-hand notation.
The Wilson line is set equal to unity by taking the light-cone gauge $A(z)\cdot z=0$.
The normalization parameter $\mu_M = m^2_M /(m_q + m_{\bar q})$ in Eq.~(\ref{Deq:1})
results from quark condensate. For the pion, $\mu_\pi = -2\la {\bar q}q\ra / f^2_\pi$ from the Gell-Mann-Oakes-Renner relation. The normalization of the DA is given by
$\int^1_0 dx \; \phi^{\cal P}_{3;M}(x) = 1$.
In order to check the existence (or absence) of the zero mode,
we again utilize the same manifestly covariant model used in the analysis of the vector meson
decay constant~\cite{CJ_V14} and then substitute the
vertex function with the more phenomenologically accessible Gaussian radial wave function
provided by our LFQM. We shall show that the analysis of the decay constants and twist-2 and twist-3 two particle DAs of pseudoscalar mesons 
 fortify
our previous conclusion drawn
from
the vector meson
case~\cite{CJ_V14}. That is, the treacherous points such as the zero-mode and the instantaneous
contributions
 present
in the covariant BS model
disappear
in the standard LFQM with
the Gaussian radial wave function but nevertheless satisfy the chiral symmetry.

\section{Manifestly Covariant Model}
\label{sec:1}
Defining the local matrix element
${\cal J}_{\cal P}
\equiv \la 0|{\bar q}i\gamma_5 q|M(P)\ra = f_M\mu_M$ of Eq.~(\ref{Deq:1})
for $z^\mu=0$,
we write the one-loop approximation  as a momentum integral
\be\label{Deq:4}
{\cal J}_{\cal P}
= N_c
\int\frac{d^4k}{(2\pi)^4} \frac{H_0} {N_p N_k}
{\rm Tr}\left[i\gamma_5\left(\slash \!\!\!p+m_q \right)\gamma_5
 \left(-\slash \!\!\!k + m_{\bar q} \right) \right],
\ee
where $N_c$ denotes the number of colors.
The denominators $N_p (= p^2 -m^2_q +i\varepsilon)$
and $N_k(= k^2 - m^2_{\bar q}+i\varepsilon)$ come from the quark propagators
of mass $m_q$ and $m_{\bar q}$ carrying the internal four-momenta $p =P -k$ and $k$, respectively.
In order to regularize the covariant loop,
we use the usual multipole ansatz $H_0=g/ N_\Lambda^2$
for the $q{\bar q}$ bound-state vertex function of a meson,
where $N_\Lambda  = p^2 - \Lambda^2 +i\varepsilon$, and $g$ and $\Lambda$ are constant parameters.
After a  little manipulation, we obtain the manifestly covariant result for
${\cal J}_{\cal P}$ as follows
\bea\label{Deq:7}
{\cal J}_{\cal P}^{\rm cov} &=& \frac{N_c g}{4\pi^2} \int^1_0 dx\int^{1-x}_0 dy (1-x-y)
\biggl\{ \frac{y(1-y)m^2_M + m_q m_{\bar q}}{C^2_{\rm cov}}
- \frac{2}{C_{\rm cov}} \biggr\},
\eea
where
$C_{\rm cov} = y(1-y) m^2_M - x m^2_q - y m^2_{\bar q} - (1-x-y) \Lambda^2$.

For the LF calculation in parallel with the manifestly covariant one,
we separate the trace term
into the on-mass-shell propagating part $[{\rm Tr}]_{\rm on}$ and the off-mass-shell instantaneous part $[{\rm Tr}]_{\rm inst}$ via $\slash\!\!\!q=\slash\!\!\!q_{\rm on} + \frac{1}{2}\gamma^+(q^- - q^-_{\rm on})$
as ${\rm Tr} = [{\rm Tr}]_{\rm on} + [{\rm Tr}]_{\rm inst}$.
We also take the reference frame where ${\bf P}_\perp =0$, i.e.,
$P=( P^+, M^2/P^+, 0)$. In this case, the LF energies of the on-mass-shell
quark and antiquark are given by
$ p^-_{\rm on} = ({\bf k}^2_\perp + m^2_q)/ xP^+$ and
$ k^-_{\rm on} = ({\bf k}^2_\perp + m^2_{\bar q})/(1-x) P^+$, respectively,
where $x=p^+/P^+$ is the LF longitudinal momentum fraction of the quark.

By the integration over $k^-$ in Eq.~(\ref{Deq:4}) and closing the contour in the lower
half of the complex $k^-$ plane, one picks up the residue at $k^-=k^-_{\rm on}$
in the region $0< k^+ < P^+$ (or $0 < x < 1$).
For the purpose of analyzing the zero-mode contribution to the decay amplitude ${\cal J}_{\cal P}$,
we denote $[{\cal J}_P]^{\rm LF}_{\rm val}$ (meaning the valence contribution to ${\cal J}_{\cal P}$)
when ${\cal J}_{\cal P}$ is obtained for $k^-=k^-_{\rm on}$ in the region of $0<x<1$. Explicitly, it is
given by
\be\label{Deq:10}
 [{\cal J}_{\cal P}]^{\rm LF}_{\rm val} = \frac{N_c}{16\pi^3}\int^{1}_0
 \frac{dx}{(1-x)} \int d^2{\bf k}_\perp
 \chi(x,{\bf k}_\perp) [{\rm Tr}]_{\rm val},
\ee
where
\be\label{Deq:11}
\chi(x,{\bf k}_\perp) = \frac{g}{[x (m_M^2 -M^2_0)][x (m_M^2 - M^2_{\Lambda})]^2},
\ee
and
$M^2_{0(\Lambda)} = [{\bf k}^{2}_\perp + m^2_q(\Lambda^2)]/x
 + ({\bf k}^{2}_\perp + m^2_{\bar q}) / (1-x)$.
The trace term for the valence contribution, i.e.
$[{\rm Tr}]_{\rm val}=[{\rm Tr}]_{\rm on} + 2 k^+ (p^--p^-_{\rm on})$, is given by
$[{\rm Tr}]_{\rm val}=2 [ M^2_0 - (m_q - m_{\bar q})^2 + (1-x) (m^2_M - M^2_0) ]$.
We find numerically that $[{\cal J}_{\cal P}]^{\rm LF}_{\rm val}$ is not equal to the
manifestly covariant result ${\cal J}_{\cal P}^{\rm cov}$ in Eq.~(\ref{Deq:7}). This
reveals
that the
decay amplitude ${\cal J}_{\cal P}$ receives a LF zero-mode contribution.
The LF zero-mode contribution to ${\cal J}_{\cal P}$ comes from the singular
$p^-$ (or equivalently $1/x$) term in the trace in the limit
of $x\to 0$ when $p^-=p^-_{\rm on}$, i.e.
$\lim_{x\to 0}{\rm Tr}(p^-=p^-_{\rm on})= 2p^-$.
Analogous to that derived in the previous analyses of weak transition form factor
calculations~\cite{Jaus99,CJ_Bc}, we identified the zero-mode operator $[{\rm Tr}]_{\rm Z.M.}$
corresponding to the zero-mode contribution $2p^-$ as follows~\cite{CJ_V14}
\be\label{Deq:15}
[{\rm Tr}]_{\rm Z.M.} = 2 (-Z_2),
\ee
where
$Z_2 = x(m_M^2 - M^2_0) + m^2_q - m^2_{\bar q} + (1-2x)m_M^2$.
Note here that there is no momentum transfer $q$ dependence for the calculation of DAs
unlike the calculation of form factors.
This zero-mode operator $[{\rm Tr}]_{\rm Z.M.}$ can be effectively included
in the valence region
and thus
\bea\label{Deq:16}
 [{\cal J}_{\cal P}]^{\rm LF}_{\rm full}
 &=& \frac{N_c}{16 \pi^3}\int^{1}_0
 \frac{dx}{(1-x)} \int d^2{\bf k}_\perp
 \chi(x,{\bf k}_\perp)[{\rm Tr}]_{\rm full},
\eea
where
$[{\rm Tr}]_{\rm full}= [{\rm Tr}]_{\rm val} + [{\rm Tr}]_{\rm Z.M.}
= 4 [ xM^2_0 + m_q (m_{\bar q}-m_q) ]$.
It can be checked that Eq.~(\ref{Deq:16}) is identical to the manifestly covariant result of
Eq.~(\ref{Deq:7}).

\section{Application to Standard Light-Front Quark Model}
\label{sec:2}
In the standard LFQM~\cite{CJ99,Jaus91}, the constituent quark and
antiquark in a bound state are required to be on-mass-shell, which is different from the covariant
BS formalism in which the constituents are off-mass-shell. In the SLF approach used in
the LFQM, the momentum space meson wave function is given by
$\Psi^{SS_z}_{\lam_1\lam_2}(x,{\bf k}_{\perp})
={\cal R}^{SS_z}_{\lam_1\lam_2}(x,{\bf k}_{\perp})
\phi_R(x,{\bf k}_{\perp})$,
where $\phi_R$ is the radial wave function and
${\cal R}^{SS_z}_{\lam_1\lam_2}$ is the
spin-orbit wave function that is obtained by the interaction-independent
Melosh transformation from the ordinary
spin-orbit wave function assigned by the quantum numbers
$J^{PC}$. The common feature of the standard LFQM
is to use the sum (i.e. invariant mass $M_0$) of the LF energy of the constituent quark
and antiquark for the meson mass in ${\cal R}^{SS_z}_{\lam_1\lam_2}$.
The virtue of using $M_0$
is to satisfy the normalization of ${\cal R}_{\lam_1 \lam_2}^{SS_z}$ automatically regardless of
any kinds of vector mesons, i.e.
$\sum_{\lam_1\lam_2}{\cal R}_{\lam_1 \lam_2}^{SS_z\dagger}{\cal R}_{\lam_1 \lam_2}^{SS_z}=1$.
The covariant forms of ${\cal R}_{\lam_1 \lam_2}^{SS_z}$ for pseudoscalar and vector mesons
are given in~\cite{CJ_V14}.
For the radial wave function $\phi_R$, we use the Gaussian wave function
\be\label{phiR}
\phi_R(x,{\bf k}_{\perp})=
(4\pi^{3/4}/\beta^{3/2}) \sqrt{\partial
k_z/\partial x} \;{\rm exp}(-{\vec k}^2/2\beta^2),
\ee
where $\beta$ is the variational parameter
fixed by the analysis of meson mass spectra~\cite{CJ_Bc,CJ99} and
$\partial k_z/\partial x$ is the Jacobian of the variable transformation
$\{x,{\bf k}_\perp\}\to {\vec k}=({\bf k}_\perp, k_z)$.

In our previous analysis of the decay constant and the twist-2 and twist-3 DAs of
a vector meson~\cite{CJ_V14},
we have shown that the SLF results in the standard LFQM is obtained by
replacement of the LF vertex function $\chi$ in BS model with the Gaussian wave function
$\phi_R$ as follows~(see Eq.~(49~)~in~\cite{CJ_V14})
\be\label{Eeq:12}
 \sqrt{2N_c} \frac{ \chi(x,{\bf k}_\perp) } {1-x}
 \to \frac{\phi_R (x,{\bf k}_\perp) }
 {\sqrt{{\bf k}^2_\perp + {\cal A}^2}}, \; m_M \to M_0,
 \ee
where ${\cal A}=(1-x)m_q + x m_{\bar q}$.
Note that the physical mass $m_M$ included in the integrand of BS
amplitude has to be replaced with the invariant mass $M_0$ ($m_M\to M_0$)
since the SLF results in the standard LFQM
are obtained from the requirement of all constituents being on their respective mass shell.
The correspondence in Eq.~(\ref{Eeq:12}) is valid again in this analysis of a pseudoscalar meson.
Applying the correspondence given by Eq.~(\ref{Eeq:12}) to the decay amplitude
${\cal J}_{\cal P}(=f_M\mu_M)$ for pseudoscalar channel given by Eq.~(\ref{Deq:16}),
we obtain the corresponding LFQM result:
\be\label{QM10}
 [{\cal J}_{\cal P}]^{\rm SLF}_{\rm full}
 = \frac{\sqrt{2N_c}}{2\cdot 16 \pi^3}\int^{1}_0 dx
 \int d^2{\bf k}_\perp
 \frac{\phi_R(x,{\bf k}_\perp)}{\sqrt{{\bf k}^2_\perp + {\cal A}^2}}
 [{\rm Tr}]_{\rm full}.
\ee
Interestingly enough we also found
the equality $[{\cal J}_{\cal P}]^{\rm SLF}_{\rm full}=[{\cal J}_{\cal P}]^{\rm SLF}_{\rm on}$
due to the fact that only the even term in $[{\rm Tr}]$ with respect to $x$
survives in the SU(2) symmetry limit~($m=m_q=m_{\bar q}$) since the Gaussian wave function $\phi_R$
and other prefactor ${\sqrt{{\bf k}^2_\perp + m^2}}$ are even in $x$.
Decomposing the trace term $[{\rm Tr}]_{\rm full} = 4 x M^2_0 = [2 + 2(2x-1)]M^2_0$ in SU(2) symmetry
limit, one can easily find that the nonvanishing contribution from $[{\rm Tr}]_{\rm full}$ is exactly the
same as $[{\rm Tr}]_{\rm on}=2M^2_0$.

Because the matrix element $[{\cal J}_{\cal P}]$ is directly related with the twist-3 DA $\phi^{\cal P}_{3;M}(x)$,
our finding in the SU(2) symmetry limit applies also for getting the correct $\phi^{\cal P}_{3;M}(x)$,
i.e. only the solution obtained from $[{\cal J}_{\cal P}]^{\rm SLF}_{\rm on}$ gives the correct $\phi^{\cal P}_{3;M}(x)$
in our LFQM, which is given by
\bea\label{QM11}
\phi^{\cal P}_{3;M}(x)
&=& -\frac{\sqrt{2N_c}}{f_M\mu_M \cdot 16 \pi^3}
\int d^2{\bf k}_\perp
\frac{\phi_R(x,{\bf k}_\perp)}{\sqrt{{\bf k}^2_\perp + {\cal A}^2}}
[M^2_0 -(m_q - m_{\bar q})^2],
\eea
where we remind that $\mu_\pi = -2\la {\bar q}q\ra / f^2_\pi$ for the pion $(m=m_q=m_{\bar q})$ case.

\section{Numerical Results}
\label{sec:3}
In our numerical calculations within the standard LFQM, we use two sets of the model parameters
(i.e. constituent quark masses $m_q$ and the gaussian parameters $\beta_{q{\bar q}}$) for the
linear and harmonic oscillator~(HO) confining potentials, which were obtained from the calculation of meson mass spectra using the variational principle in our LFQM~\cite{CJ_Bc,CJ99}.
Since the numerical results of the twist-2 DAs of pseudoscalar mesons were given in~\cite{CJ_DA},
we shall focus
on the calculation of the twist-3 DAs of $\pi$ an $K$ mesons in this section.

Our LFQM predictions for the decay constants of $\pi$ and $K$ mesons obtained from the linear~[HO] potential
parameters are $f^{\rm SLF}_\pi=130 ~[131]$ MeV and $f^{\rm SLF}_K= 161~ [155]$ MeV, which are in good agreement with the experimental data~\cite{PDG14}; $f^{\rm Exp.}_\pi=(130.41\pm 0.03 \pm 0.20)$ MeV and
$f^{\rm Exp.}_K =(156.2 \pm 0.3 \pm 0.6 \pm 0.3)$ MeV.
We then obtain the quark condensate $\la q{\bar q}\ra$, which enters the normalization
of twist-3 pion DA $\phi^{\cal P}_{3;\pi}(x)$ given by Eq.~(\ref{QM11}), as
$-(285.8 {\rm MeV})^3 [-(263.7 {\rm MeV})^3]$ for the linear~[HO] potential parameters.
Our LFQM results, especially the one obtained from HO parameters, are quite comparable with the commonly
used phenomenological value $\la {\bar q}q\ra = -(250 {\rm MeV})^3$.

\begin{figure}
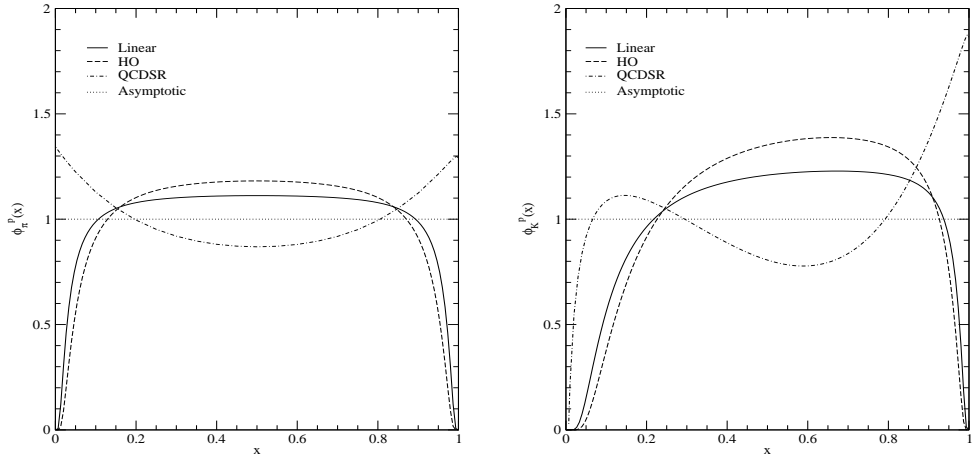

\vspace{0.8cm}
\centering
\includegraphics[height=6cm, width=6cm]{fig1a.eps}
\hspace{0.5cm}
\includegraphics[height=6cm, width=6cm]{fig1b.eps}
\caption{ The twist-3 DAs $\phi^{\cal P}_{3;M}(x)$ for $\pi$ (left panel) and
$K$ (right panel) mesons obtained from the linear~(solid line) and HO (dashed line) parameters
compared with the QCD sum rule result (dot-dashed line)~\cite{BBL} as well as the asymptotic one
(dotted line)~\cite{BF}.}
\label{fig1}
\end{figure}

In Fig.~\ref{fig1}, we show the twist-3 DAs $\phi^{\cal P}_{3;M}(x)$ [see Eq.~(\ref{QM11})],
which is free from the explicit instantaneous as well as zero-mode contributions,
for $\pi$ (left panel) and
$K$ (right panel) mesons obtained from the linear~(solid line) and HO (dashed line) parameters.
We also compare our results with the the asymptotic DA $[\phi^{\cal P}_{3;M}]_{\rm as}(x)=1$
(dotted line)~\cite{BF} as well as the QCD sum rule (QCDSR) prediction (dot-dashed line)~\cite{BBL}.
For the pion case, our results obtained from both model parameters
not only show the symmetric forms anticipated from the
isospin symmetry but also reproduce the exact asymptotic result
$[\phi^{\cal P}_{3;\pi}]_{\rm as}(x)=1$ in the chiral symmetry ($m_q\to 0$) limit.
Our LFQM result of $\phi^{\cal P}_{3;\pi}(x)\to [\phi^{\cal P}_{3;\pi}]_{\rm as}(x)$
in the chiral symmetry limit is consistent with the conclusion drawn from our previous
analysis~\cite{CJ_V14} of the twist-2 ($\phi^{||}_{2;\rho}(x)$)
and twist-3 ($\phi^{\perp}_{3;\rho}(x)$) $\rho$ meson DAs, where both DAs also
remarkably reproduce the exact asymptotic DAs in the chiral symmetry limit.
This example shows again that our LFQM prediction satisfies the chiral symmetry consistent
with the QCD if one correctly implement the zero-mode link to the QCD vacuum.
It is also interesting to note that while our results of $\phi^{\cal P}_{3;\pi}(x)$ become zero at the
end points of $x$, the result of Ref.~\cite{BBL} obtained from using the current quark
masses does not vanish at the end points.
For $K$ meson case, we assign the momentum fractions $x$ for $s$-quark
and $(1-x)$ for the light $u(d)$-quark. Due to the flavor SU(3) symmetry breaking effect,
$\phi^{\cal P}_{3;K}(x)$ obtained from both model parameters are asymmetric and the peak points
are moved to the right of the $x=0.5$ point indicating that the $s$-quark carries more longitudinal
momentum fraction than the light $u(d)$-quark. As in the case of pion, our results
of $\phi^{\cal P}_{3;K}(x)$ are very different from that of Ref.~\cite{BBL}.
The twist-3 quark DA can be expanded in terms of the Gegenbauer polynomials
$C^{1/2}_n$ as
$\phi^{\cal P}_{3;M} =
 [\phi^{\cal P}_{3;M}]_{\rm as}(x)
 \biggl[ 1 + \sum^{\infty}_{n=1} a^{{\cal P}}_{n,M}C^{1/2}_n(2x-1) \biggr]$,
where $[\phi^{\cal P}_{3;M}]_{\rm as}(x)=1$. The coefficients
$a^{{\cal P}}_{n,M}$ are called the Gegenbauer moments and can be obtained by
$a^{\cal P}_{n,M}(x) = (2n+1) \int^1_0 dx C^{1/2}_n(2x-1)\phi^{\cal P}_{3;M}(x)$.

\begin{table}[t]
\caption{The Gegenbauer moments of twist-3 $\phi^{\cal P}_{3;\pi}$ for pion obtained from the
linear and HO potential models compared other model estimates. }
\centering
\label{t1}
\begin{tabular}{llll}
\hline\noalign{\smallskip}
Models & $a^{\cal P}_{2,\pi}$ & $a^{\cal P}_{4,\pi}$ & $a^{\cal P}_{6,\pi}$ \\[3pt]
\tableheadseprule\noalign{\smallskip}
HO & -0.5816 &  -0.4110 &  -0.1725 \\
Linear & -0.3979 &  -0.3739 &  -0.2500  \\
SR~\cite{BBL}  & 0.4373 &  -0.0715 &  -0.1969  \\
$\chi$QM~\cite{NK06}  & -0.4307 &  -0.5559 &  -0.1784  \\
\noalign{\smallskip}\hline
\end{tabular}
\end{table}

\begin{table}[t]
\caption{The Gegenbauer moments of twist-3 $\phi^{\cal P}_{3;K}$ for kaon obtained from the
linear and HO potential models compared other model estimates. }
\centering
\label{t2}
\begin{tabular}{lllllll}
\hline\noalign{\smallskip}
Models &  $a^{\cal P}_{1,K}$ & $a^{\cal P}_{2,K}$ & $a^{\cal P}_{3,K}$
& $a^{\cal P}_{4,K}$ & $a^{\cal P}_{5,K}$ & $a^{\cal P}_{6,K}$ \\
\tableheadseprule\noalign{\smallskip}
HO & 0.3187 &  -0.7800 &  -0.0647 & -0.2923 & -0.2223 & -0.0396\\
Linear & 0.2662 &  -0.6104 &  0.0486 & -0.3361 &  -0.1454 & -0.1161 \\
SR~\cite{BBL}&  0.1837 &  0.2707 &  0.3953 & -0.2469 &  0.0550 & -0.2436\\
$\chi$QM~\cite{NK06}  & 0.0236 &  -0.6468 &  -0.0367 & -0.3724 &  -0.0200 & -0.0940 \\
\noalign{\smallskip}\hline
\end{tabular}
\end{table}
In Tables~\ref{t1} and~\ref{t2}, we list the calculated Gegenbauer moments of twist-3 pion and
kaon DAs, respectively, obtained from the linear and HO potential models at the scale $\mu\sim 1$ GeV.
We also compare our results with other model estimates calculated
at the scale $\mu=1$ GeV, e.g. QCD sum rules~\cite{BBL} and the chiral quark model ($\chi$QM)~\cite{NK06}.
For the pion case, as expected from the isospin symmetry, all odd Gegenbauer moments are all zero.
While our results are quite different from those of QCD sum rules~\cite{BBL},
they are consistent with the $\chi$QM predictions~\cite{NK06}.
For the kaon case, the odd moments are nonzero due to the flavor SU(3) symmetry
breaking effects and the first moments $a^{\cal P}_{1,K}$ is proportional to the difference between
the longitudinal momenta of the strange and nonstrange quark in the two-particle Fock component.
Our results for the twist-3 DA are overall in good agreement with those of $\chi$QM~\cite{NK06}
but quite different from QCD sum rule estimates~\cite{BBL}. We also should note that the very
small value of $a^{\cal P}_{1,K}$ obtained from $\chi$QM~\cite{NK06} is due to the more symmetric and
flat shape of DA than ours.

\section{Summary}
\label{sec:4}

In this paper, following our previous work~\cite{CJ_V14}, we have discussed a
wave function dependence of the LF zero-mode contributions to the twist-3 two-particle
DA $\phi^{\cal P}_{3;M}$ of a pseudoscalar meson between the two models, i.e. the exactly solvable manifestly
covariant BS model and the more phenomenologically accessible realistic LFQM using the
SLF approach. Linking the covariant BS model to the standard LFQM,
we found that the
correspondence relation given by Eq.~(\ref{Eeq:12}) between the two models
applies not only for the vector meson~\cite{CJ_V14} but also for the pseudoscalar meson.
The remarkable finding in linking the covariant BS model to the standard LFQM
is that the treacherous points such as the zero-mode contributions and the instantaneous
ones existed in the covariant BS model become absent in the LFQM with the Gaussian wave function.
Our numerical results of $\phi^{\cal P}_{3;\pi}$ not only show the symmetric forms
anticipated from the isospin symmetry but also reproduce the exact asymptotic result
$[\phi^{\cal P}_{3;\pi}]_{\rm as}(x)=1$ in the chiral symmetry ($m_q\to 0$) limit.
For the kaon case, the results of $\phi^{\cal P}_{3;K}$ show asymmetric form because of the
flavor SU(3) symmetry breaking effects. Our results for the Gegenbauer moments of
twist-3 pion and kaon DAs are overall in good agreement with the chiral quark model~\cite{NK06} but
quite different from those of QCD sum rule estimates~\cite{BBL}.
Since we have shown that the absence/existence of the zero-mode contribution as well as the instantaneous contribution to the twist-3 DA may depend on the used bound state vertex function, it would be interesting to study this process with other vertex function such the symmetric product ansatz suggested in Eq. (38) of Ref.~\cite{FPPS}.



\end{document}